# Least quartic Regression Criterion with Application to Finance

GIUSEPPE ARBIA

*Department of Statistics, Università Cattolica del Sacro Cuore, Rome, Italy (giuseppe.arbia@rm.unicatt.it)*

____________________________________________________________________

**Abstract**

This article proposes a new method for the estimation of the parameters of a simple linear regression model which accounts for the role of co-moments in non-Gaussian distributions being based on the minimization of a quartic loss function. Although the proposed method is very general, we examine its application to finance. In fact, in this field the contribution of the co-moments in explaining the return-generating process is of paramount importance when evaluating the systematic risk of an asset within the framework of the Capital Asset Pricing Model (*CAPM*). The suggested new method contributes to this literature by showing that, in the presence of non-normality, the regression slope can be expressed as a function of the co-kurtosis between the returns of a risky asset and the market proxy. The paper provides an illustration of the method based on some empirical financial data referring to 40 industrial sector assets' rates of the Italian stock market.

____________________________________________________________________

# 1. INTRODUCTION

Traditional linear regression models based on the normality assumption neglect any role to the higher moments of the underlying distribution. This approach is not justified in many situations where the phenomenon is characterized by strong non-normalities like outliers, multimodality, skewness and kurtosis. Although this situations may occur in many circumstances, quantitative finance is a field where the consequences of non-normalities are particularly relevant and may affect dramatically the investors' decisions. Consider, for instance, the capital asset pricing model (henceforth *CAPM*; see Sharpe, 1964 and Lintner, 1965) a well-known equilibrium model which assumes that investors compose their portfolio on the basis of a trade-off between the expected return and the variance of the return of the entire portfolio. In the traditional *CAPM* framework a regression slope indicates how strong fluctuations in the returns of a single asset are related to movements of the market as a whole. As such it can be interpreted as a measure of the market (often called "systematic") risk associated to that asset. Since it is impossible to completely eliminate this risk with diversification, investors are compensated with a risk premium. In its original formulation based on Least Squares estimation, *CAPM* is restricted to the first two moments of the empirical distribution of past returns, however, due to the complexity of the return generating process in future markets, this simple model often fails in detecting risk premia. Many papers in the financial literature heavily criticized such an approach emphasizing the role of co-moments in investors' decisions and leading to a series of modifications that incorporate the consideration of higher moments of the distribution of returns (e. g. Barone-Adesi et al. , 2004).

This article moves along these lines. In particular the originality of our contribution consists in defining a linear regression estimation criterion which is based on a quartic loss function in place of the usual quadratic specification. Although the proposed methodology is very general, in the present paper we examine in particular its application to finance. Using the quartic loss function criterion, in fact, we show that it is possible within the CAPM framework, to take into account higher moments of the assets returns and their co-moments with the market portfolio, and also to provide an index of each asset's systematic risk which accounts for non-normalities by incorporating higher-order.

To achieve these aims the article is organized as follows. Section 2 is devoted to a thorough literature review on the role of higher moments in financial systematic risk evaluation. This background is useful not only to motivate the interest in the new approach, but also to provide the reader with the framework within which the results of the proposed methodology could be interpreted. In Section 3.1 we formally introduce various definitions of co-skewness and co-kurtosis and we clarify their statistical nature with particular reference to the analysis of systematic risk. Section 3.2 is devoted to the introduction of a regression estimator which involves the minimization of the fourth power of the regression errors. In Section 4 we illustrate the use of the proposed methodology using some real data referring to the returns of a set of industrial companies of the Italian stock market. Finally Section 5 concludes and traces some possible directions for future researches in this field.

## 2. BACKGROUND: THE ROLE OF HIGHER MOMENTS IN FINANCIAL RISK MEASUREMENT

The traditional theory of CAPM uses essentially a Least Square linear regression strategy to measure the relationship between the fluctuation of the single asset in relation to the market thus quantifying a measure of the *systematic* risk of the asset. In this respect only the first two moments of the joint distribution between the asset and the market are relevant in the analysis (Sharpe, 1964; Lintner, 1965; Mossin, 1966). In the last decades many papers have recognized the shortcomings associated to such an approach (starting from the criticism contained, e. g., in Fama and French, 1992), and extended it so as to incorporate considerations linked to the higher-order moments of the generating probability distributions. In particular some authors (like, e. g. Ranaldo and Favre, 2005) agree on the fact that the use of a pricing model limited to the first two moments may be misleading and may wrongly indicate insufficient compensation for the investment. They thus indicate that a higher-moment approach is more appropriate to detect non-linear relationships between assets return and the portfolio returns while accommodating for the specific risk-return payoffs. The financial literature is very rich of contributions that include considerations related to third moment. For instance papers like Rubinstein (1973), Kraus and Litzenberger (1976; 1983), Barone-Adesi and Talwar (1983), Barone-Adesi (1985), Sears and Wei (1988) and Harvey and Siddique (2000) propose a three-moment CAPM which includes the third-order moment. More recently papers like Barberis and Huang (2008), Brunnermeier et al. (2007), Mitton and Vorkink (2007), Boyer et al. (2010) and Green and Hwang (2012) empirically document that the skewness of individual assets may have an influence on portfolio decisions. Rather surprisingly, comparatively less attention has been paid to the role of the fourth moment. In this respect Krämer and Runde (2000) show evidences of leptokurtic stock returns, while, more recently, Conrad et al. (2013) discuss the role of (risk neutral) skewness and kurtosis showing that they are strongly related to future returns.

Co-moments of the asset's distribution with the market portfolio have also been recognized as influencing investors' expected returns. In this respect Hwang and Satchell (1999), Harvey and Siddique (2000; 2002) and Bakshi et al. (2003) introduce the idea of co-moments analysis in financial market risk evaluation while Dittmar (2002) tests explicitly the influence of co-skewness and co-kurtosis on investors' decisions, showing that systematic kurtosis is better than systematic skewness in explaining market returns. Fang and Lai (1997) and Christie-David and Chaudry (2001) describe a model where the excess returns are expressed as a function of the covariance, of the co-skewness and of the co-kurtosis between the returns of a risky asset and those of the investor's portfolio. They show that volatility is an insufficient measure of risk for risk-averse agents and that systematic skewness and systematic kurtosis increase the explanatory power of the return generating process of future markets.[1] Their empirical results agree on the fact that investors are adverse both to variance and kurtosis in their portfolio requiring higher excess rates as a compensation. The papers by Fang and Lai (1997) and Christie-David and

---

[1] Further references to the use of higher moments in analyzing future markets may be found in Levy (1969), Bandrinath and Cahtterjee (1988), Hudson et al. (1987) and Scott and Horvat (1980) Jurcenzko and Maillet, (2002) Hwang and Satchell (1999) amongst the others. Other approaches related to higher order considerations in the evaluation of risk may be found in the extreme value approach suggested by McNeil and Frey (2000), in the idea of bivariate VaR by Arbia (2003) and in the tail VaR introduced by Bargès *et al.* (2009).

Chaudry (2001) have the merit of having introduced in a formal way the consideration of co-skwness and co-kurtosis in a CAPM framework showing their empirical relevance in explaining risk premia. In this paper we follow the same approach showing how higher moments considerations can be formally taken into account redefining the estimation criterion of the fundamental systematic risk regression.

## 3. METHODOLOGY

3.1 Co-moments

In this section we will introduce some formal definitions of co-skewness and co-kurtosis that will be used later while discussing an augmented version of the *CAPM* which takes into account higher moments of the distribution of the returns. To start with, if we consider two random variables *X* and *Y*, let us define the generic bivariate moment of order *r,s* centered around the mean as $\mu_{r,s} = E(x^r y^s)$ where $x = X - E(X)$ and $y = Y - E(Y)$. Similarly, the standardized bivariate moment of order *r,s* centered around the mean is defined as $\lambda_{r,s} = \dfrac{\mu_{r,s}}{\sigma_x^r \sigma_y^s}$ with $\sigma_x = \sqrt{\mu_{2,0}}$ and analogously for $\sigma_y$. Starting from these definitions, the co-skewness of a bivariate distribution is defined formally as the mixed moments of orders *r* and *s* such that *r+s=3*, that is $\mu_{1,2}$ and $\mu_{2,1}$. In a bivariate normal distribution $\mu_{1,2} = \mu_{2,1} = 0$ (see Kendall and Stuart, 1983 and Kotz *et al.*, 2000) however, in general non-normal distributions, we have, conversely, $\mu_{1,2} \in \Re$ and $\mu_{2,1} \in \Re$. The corresponding standardized moment is defined as $\lambda_{2,1} = \frac{\mu_{2,1}}{\sigma_x^2 \sigma_y}$ and similarly for $\lambda_{1,2}$. Figure 1 illustrates graphically the behavior of positive and negative co-skewness showing the two density functions and the associated scatter diagrams.

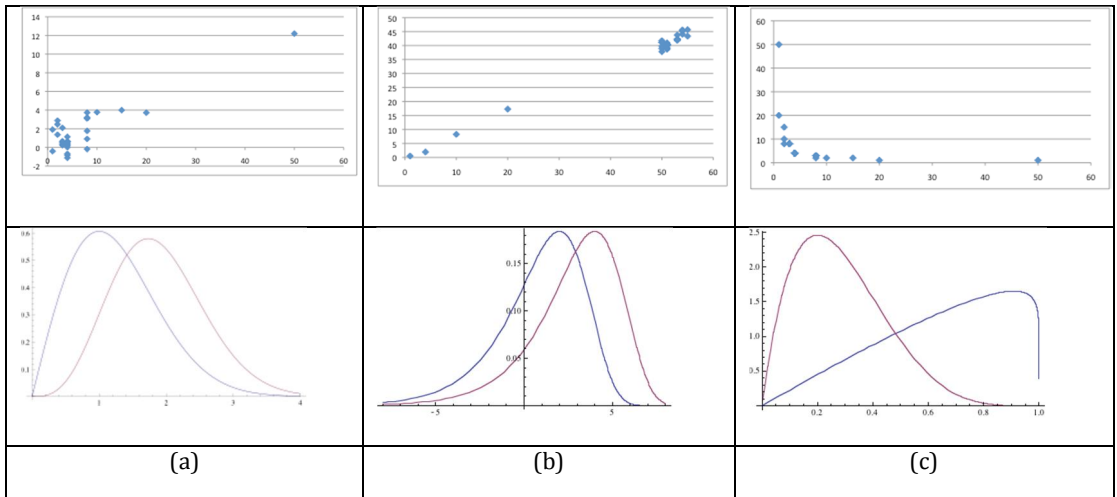

*Figure 1: Various typologies of co-skewness: (a) positive left-tail co-skewness; (b) positive right-tail co-skewness; (c) negative co-skewness.* Scatter diagram (upper pane). Probability Density Function (lower panel).

In finance for the systematic co-skewness of a joint distribution, a slightly different version of the standardized moment is usually considered which is defined by the term $\frac{\mu_{2,1}}{\mu_{3,0}}$ (Christie-David and Chaudry, 2001; Fang and Lai, 1997; Harvey and Siddique, 2000; 2002). Always referring to the financial interpretation, from Figure 1 it can be argued that, in general, investors will tend to avoid positive right-tail co-skewness (involving cases of big losses when the risk-free asset also experiences big losses) and to privilege instead left-tail positive co-skewness (involving cases of big revenues when the risk-free asset also experiences big revenues). Prudential investors will prefer negative co-skewness to mitigate the risk of losses when the markets performs badly.

In a similar fashion, the co-kurtosis of a bivariate distribution is defined by the mixed moments of orders *r* and *s* such that *r+s=4*. Co-kurtosis can therefore assume three different manifestations defined respectively by the mixed moments $\mu_{1,3}$, $\mu_{3,1}$ and $\mu_{2,2}$ whose behavior is likely to be correlated in empirical situations. In a bivariate normal distribution characterized by a correlation $\rho$, we have that $\mu_{1,3} = 3\rho\sigma_1\sigma_2^3$ and analogously for $\mu_{3,1}$ (see Kendal and Stuart, 1983; Kotz *et al.*, 2000). In non-Gaussian distributions, we have, instead, $\mu_{1,3} \in \Re$ and $\mu_{3,1} \in \Re$. The corresponding standardized moment is defined as $\lambda_{1,3} = \frac{\mu_{1,3}}{\sigma_1\sigma_2^3}$ and similarly for $\lambda_{3,1}$. A relative measure of co-skewness with respect to the bivariate normal distribution is intuitively provided by the expression $\kappa_{1,3} = \mu_{1,3} - 3\rho\sigma_1^2\sigma_2^2$ and similarly for $\kappa_{3,1}$. Analogously we can look at co-kurtosis through the mixed moment $\mu_{2,2}$. In a bivariate normal distribution with correlation $\rho$, we have that $\mu_{2,2} = \sigma_1^2\sigma_2^2(1+2\rho^2)$ (see Kendal and Stuart, 1983; Kotz *et al.*, 2000). Conversely in non-Gaussian distributions, we have, instead, $\mu_{2,2} \in \Re^+$. The corresponding standardized moment is defined as $\lambda_{2,2} = \frac{\mu_{2,2}}{\sigma_1^2\sigma_2^2}$ and the relative measure with respect to the bivariate normal distribution is provided by the quantity $\kappa_{2,2} = \mu_{2,2} - \sigma_1^2\sigma_2^2(1+2\rho^2)$. In the financial literature, the systematic co-kurtosis of a risky asset has been often defined as a modified version of the standardized *(3,1)* moment given by $\frac{\mu_{3,1}}{\mu_{4,0}}$ (Christie-David and Chaudry, 2001; Fang and Lai, 1997) while the joint moment of order (2,2) has been completely neglected. The range of possible situations that may arise here are more complex than those considered for co-skewness. Figure 2 illustrates graphically only some of the possible cases where positive and negative co-kurtosis may emerge in empirical circumstances. Again finance provides interesting substantive interpretations of the co-kortosis parameters. In fact, looking at Figure 2, one may argue that, generally speaking, investors prefer negative co-kurtosis that reduces the chances of big losses when the markets performs badly. This conclusions is in accordance with the speculations of Fang and Lai (1997) and Christie-David and Chaudry (2001) on the investors' behavior in the presence of kurtosis.

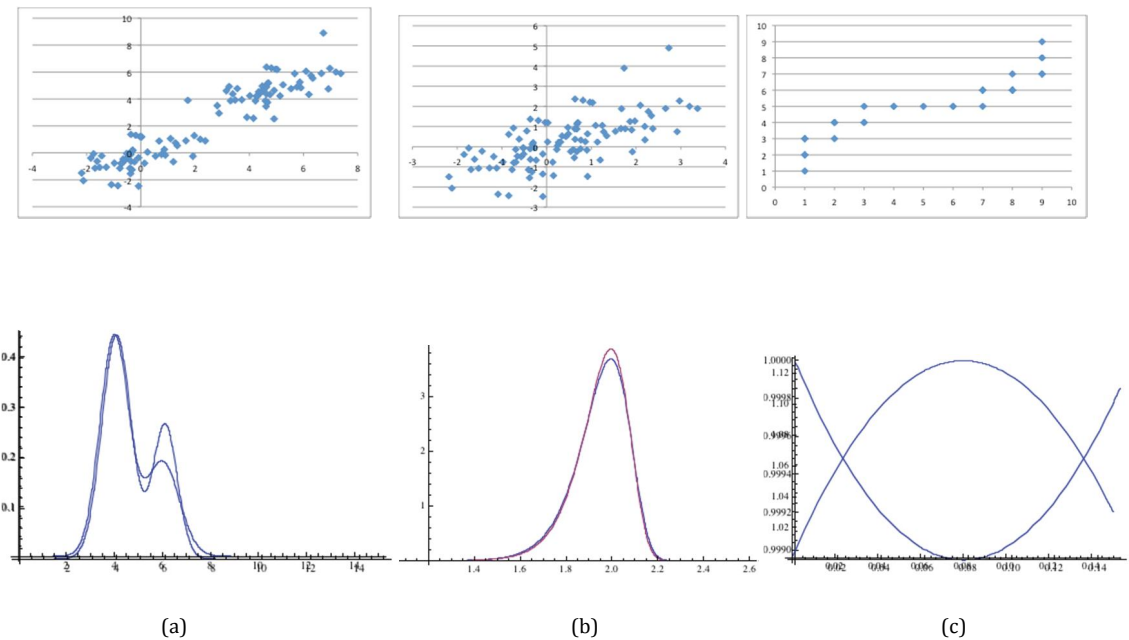

(a)  (b)  (c)

*Figure 2: Various typologies of co-skewness: (a) positive lepto co-kurtosis; (b) positive plati co-kurtosis; (c) negative co-kurtosis. Scatter diagram (upper pane). Probability Density Function (lower panel).*

3.2 The least quartic criterion

In what follows, we introduce a regression interpolation criterion that takes into account higher-order moments characteristics in non-normal situations. Alternative estimation approaches that account for the presence of non-normality and outliers in regression may be found in the contributions of Theil (1950) and Sen (1968) among the others.

Let us start considering a simple linear regression model:

$$y_i = bx_i + \varepsilon_i \tag{1}$$

with both *x* and *y* expressed in terms of deviations from their respective expected values.

To find the optimal interpolating line, in place of the familiar Least Square criterion based on a quadratic loss function, let us define a quartic loss function *l(b)*:

$$l(b) = \sum_{i=1}^{n} \varepsilon_i^4 = \sum_{i=1}^{n} (y_i - bx_i)^4 \tag{2}$$

A motivation for the choice of a *least quartic* criterion may be found in the financial literature discussed in Section 2 and in particular in the empirical findings of Fang and Lai (1997) and Christie-David and Chaudry (2001) related to investors adversity to portfolio kurtosis.

By expanding Equation (2) we have:

$$l(b) = \mu_{40}b^4 - 4b^3\mu_{31} + 6b^2\mu_{22} - 4b\mu_{13} + \mu_{04} \qquad (3)$$

In the above expression, $\mu_{40} = \sum x_i^4$ is the kurtosis of the independent variable, $\mu_{31} = \sum x_i^3 y_i$, $\mu_{13} = \sum x_i^3 y_i$, $\mu_{2,2} = \sum x_i^2 y_i^2$ represent the various measures of co-kurtosis between variables *x* and *y* described in Section 2.1, and, finally, $\mu_{04} = \sum y_i^4$ represents the kurtosis of the dependent variable. To help a visual interpretation, Figure 3 reports an example of a quartic polynomial which, in general, identifies a curve with two relative minima and one maximum. However, in the specific case we are examining, the polynomial to be minimized is subject to a series of constraints deriving from the intrinsic nature of the problem. In particular, the polynomial parameters referring to the kurtosis and to the co-kurtosis $\mu_{2,2}$ are bounded to be positive by definition.

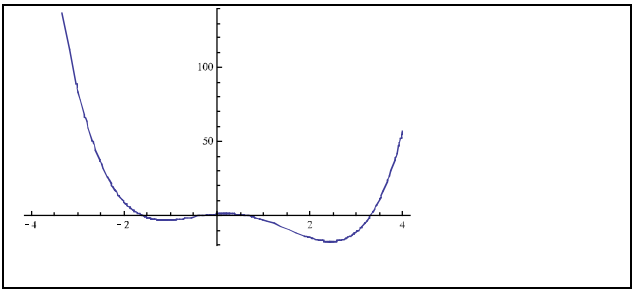

*Figure 3: Plot of a quartic polynomial. For the sake of illustration we set $\mu_{40} = 1$; $\mu_{31} = -4$; $\mu_{22} = -6$; $\mu_{13} = 2$ and $\mu_{04} = 1$*

In order to derive the minimum of the loss function, let us now set to zero the first derivative of Equation (3). We obtain:

$$\frac{\partial}{\partial b} l(b) = -4 \sum_{i=1}^{n} (y_i - bx_i)^3 x_i = 0$$

$$-\sum_{i=1}^{n} (y_i - bx_i)^3 x_i = 0$$

which leads to

$$\mu_{40} b^3 - 3\mu_{31} b^2 + 3\mu_{22} b - \mu_{13} = 0 \tag{4}$$

This equation admits two imaginary roots and one in the real field. The real solution is expressed through the following equation (Jacobson, 2009)[2]:

$$b_{LQ} = \frac{\mu_{31}}{\mu_{40}} - \left[\sqrt[3]{2}\left(-9\mu_{31}^2 + 9\mu_{22}\mu_{40}\right)\right] / \left\{3\mu_{40}\left[54\mu_{31}^3 - 81\mu_{22}\mu_{31}\mu_{40} + 27\mu_{13}\mu_{40}^2 + \sqrt{4\left(-9\mu_{31}^2 + 9\mu_{22}\mu_{40}\right)^3 + \left(54\mu_{31}^3 - 81\mu_{22}\mu_{31}\mu_{40} + 27\mu_{13}\mu_{40}^2\right)^2}\right]^{1/3}\right\} +$$

$$+ \frac{1}{3 \times 2^{1/3} \mu_{40}} \left(54\mu_{31}^3 - 81\mu_{22}\mu_{31}\mu_{40} + 27\mu_{13}\mu_{40}^2 + \sqrt{4\left(-9\mu_{31}^2 + 9\mu_{22}\mu_{40}\right)^3 + \left(54\mu_{31}^3 - 81\mu_{22}\mu_{31}\mu_{40} + 27\mu_{13}\mu_{40}^2\right)^2}\right)^{1/3} \tag{5}$$

---

[2] All calculations in this sections where run using Wolfram Mathematica ® software, version 8.

We will refer to the value thus obtained as to the Least Quartic estimator (*LQ*) of the *CAPM* regression slope. Notice that in the normal distribution case (Kendall and Stuart, 1983) we have:

$$\mu_{40} = 3\sigma_x^4; \mu_{31} = 3\rho\sigma_x^3\sigma_y; \mu_{22} = \sigma_x^2\sigma_y^2(1+2\rho^2); \mu_{13} = 3\rho\sigma_x\sigma_y^3 \text{ and } \mu_{04} = 3\sigma_y^4$$

and, following the intuition, the real solution of Equation (3), which is reported in expression (5), reduces to the ordinary least squares (*LS*) solution.

From Equation (4) we can derive the second-order condition:

$$\frac{\partial^2}{\partial b}l(b) = 3\mu_{40}b^2 - 6\mu_{3,1}b + 3\mu_{2,2}$$

and this quantity is negative, thus identifying a minimum of the loss function, only if $b < \left|\frac{\mu_{3,1}}{\mu_{4,0}}\sqrt{\mu_{3,1} - \mu_{4,0}\mu_{2,2}}\right|$. However, if we interpret the loss function as a way of finding the best fit of a linear function to a cloud of points, it admits no theoretical maximum in the real field in that we can always imagine a curve that is infinitely far from the cloud of points. So the real solution reported in Equation (5) can be interpreted as a minimum.

4. EMPIRICAL EXAMPLE

In this section the Least Quartic criterion is illustrated with reference to the daily observations of some Italian asset prices in a period of time ranging from the 1st of January 2009 to the 12nd of November 2012. This period covers a total of 1,419 days when 995 observations related to the working days were collected. Sourced from Bloomberg, we considered the prices of 36 companies classified into 8 industrial sectors (namely Utilities, Industrial, Financial, Consumer cyclical, Consumer non-cyclical, Energy, Communication and Technology). Four companies were excluded from the set of 40 assets included in the original sample (namely, ENEL Greenpower, Exor, Fiat Industrial and Salvatore Ferragamo) due to incompleteness of the data in the observed period. Table 1 provides some characteristics of the sample and some summary measures. Relatively high variability is observed for Banca Popolare di Milano, Banca Popolare Società Cooperative s.r.l. (SCARL), Mediobanca and Pirelli & C. All frequency distributions are very irregular and far from being Gaussian displaying multiple modes, asymmetry and anomalous tails. (For the sake of succinctness, in Figure 4 we only report some examples). There is a marked prevalence of negative skew (27 out of the 36 assets) and of negative excess kurtosis (32 out of the 36 assets). Almost all values of both skewness and kurtosis are significantly different from the Gaussian distribution reference values at the usual significance levels (see Table 1).

Both Kolmogorv-Smirnof and Shapiro-Wilks tests, rejects the hypothesis of normality with more than 0.1 % significance level for all assets.

Table 1: Summary statistics: daily observations of the price of the FTSE MIB index and of the price of the 36 sampled equities (Source: Bloomberg. Period: 1/1/2009-19/11/2012). High values of the Coefficient of Variation, of positive Skewness and of positive Excess kurtosis are highlighted in bold. Significance: * = 10 %; ** = 5%.

| | Name | | Country | Sector | mean | variance | st dev | cv | skewness | | excess kurtosis | |
|---|---|---|---|---|---|---|---|---|---|---|---|---|
| | FTSE MIB INDEX | FTSEMIB Index | country | industry sector | 18796,547 | 18796,547 | 10283475,346 | 0,171 | -0,242 | ** | -1,305 | ** |
| 1 | A2A SPA | A2A IM Equity | ITALY | Utilities | 1,018 | 0,102 | 0,319 | 0,313 | -0,801 | ** | -0,555 | ** |
| 2 | ANSALDO STS SPA | STS IM Equity | ITALY | Industrial | 7,221 | 1,282 | 1,132 | 0,157 | -0,072 | | -0,821 | ** |
| 3 | ASSICURAZIONI GENERALI | G IM Equity | ITALY | Financial | 14,288 | 6,512 | 2,552 | 0,179 | -0,173 | ** | -0,761 | ** |
| 4 | ATLANTIA SPA | ATL IM Equity | ITALY | Consumer, Non-cyclical | 12,756 | 3,708 | 1,925 | 0,151 | -0,264 | ** | -1,056 | ** |
| 5 | AUTOGRILL SPA | AGL IM Equity | ITALY | Consumer, Cyclical | 8,122 | 2,413 | 1,553 | 0,191 | -0,745 | ** | **0,390** | ** |
| 6 | AZIMUT HOLDING SPA | AZM IM Equity | ITALY | Financial | 7,130 | 2,252 | 1,501 | 0,210 | -0,468 | ** | -0,238 | * |
| 7 | BANCA MONTE DEI PASCHI S | BMPS IM Equity | ITALY | Financial | 0,692 | 0,108 | 0,328 | 0,474 | -0,180 | ** | -1,246 | ** |
| 8 | BANCA POPOL EMILIA ROMA | BPE IM Equity | ITALY | Financial | 7,766 | 4,339 | 2,083 | 0,268 | -0,634 | ** | -0,829 | ** |
| 9 | BANCA POPOLARE DI MILAN | PMI IM Equity | ITALY | Financial | 0,947 | 0,229 | 0,479 | **0,505** | **0,115** | * | -1,384 | ** |
| 10 | BANCO POPOLARE SCARL | BP IM Equity | ITALY | Financial | 2,510 | 1,521 | 1,233 | **0,491** | **0,205** | ** | -1,298 | ** |
| 11 | BUZZI UNICEM SPA | BZU IM Equity | ITALY | Industrial | 8,985 | 2,568 | 1,602 | 0,178 | **0,229** | ** | -0,576 | ** |
| 12 | DAVIDE CAMPARI-MILANO S | CPR IM Equity | ITALY | Consumer, Non-cyclical | 4,388 | 1,264 | 1,124 | 0,256 | -0,419 | ** | -0,827 | ** |
| 13 | DIASORIN SPA | DIA IM Equity | ITALY | Consumer, Non-cyclical | 25,498 | 31,270 | 5,592 | 0,219 | -0,136 | ** | -1,010 | ** |
| 14 | ENEL SPA | ENEL IM Equity | ITALY | Utilities | 3,600 | 0,392 | 0,626 | 0,174 | -0,447 | ** | -0,656 | ** |
| 15 | ENI SPA | ENI IM Equity | ITALY | Energy | 16,479 | 1,419 | 1,191 | 0,072 | -1,058 | ** | **1,322** | ** |
| 16 | FIAT SPA | F IM Equity | ITALY | Consumer, Cyclical | 4,444 | 1,856 | 1,362 | 0,307 | **0,620** | ** | **0,223** | * |
| 17 | FINMECCANICA SPA | FNC IM Equity | ITALY | Industrial | 7,705 | 9,263 | 3,043 | **0,395** | -0,414 | ** | -1,295 | ** |
| 18 | IMPREGILO SPA | IPG IM Equity | ITALY | Industrial | 2,403 | 0,184 | 0,429 | 0,179 | **0,593** | ** | -0,374 | ** |
| 19 | INTESA SANPAOLO | ISP IM Equity | ITALY | Financial | 1,923 | 0,344 | 0,586 | 0,305 | -0,079 | | -1,208 | ** |
| 20 | LOTTOMATICA GROUP SPA | LTO IM Equity | ITALY | Consumer, Cyclical | 13,375 | 3,857 | 1,964 | 0,147 | -0,015 | | -0,686 | ** |
| 21 | LUXOTTICA GROUP SPA | LUX IM Equity | ITALY | Consumer, Non-cyclical | 20,698 | 21,178 | 4,602 | 0,222 | -0,120 | * | -0,243 | * |
| 22 | MEDIASET SPA | MS IM Equity | ITALY | Communications | 3,707 | 2,090 | 1,446 | **0,390** | -0,260 | ** | -1,174 | ** |
| 23 | MEDIOBANCA SPA | MB IM Equity | ITALY | Financial | 6,445 | 3,017 | 1,737 | 0,270 | -0,403 | ** | -0,679 | ** |
| 24 | MEDIOLANUM SPA | MED IM Equity | ITALY | Financial | 3,415 | 0,379 | 0,615 | 0,180 | **0,216** | ** | -0,581 | ** |
| 25 | PARMALAT SPA | PLT IM Equity | ITALY | Consumer, Non-cyclical | 1,826 | 0,093 | 0,305 | 0,167 | **0,619** | ** | **0,521** | ** |
| 26 | PIRELLI & C. | PC IM Equity | ITALY | Consumer, Cyclical | 5,619 | 4,006 | 2,001 | **0,356** | -0,008 | | -0,857 | ** |
| 27 | PRYSMIAN SPA | PRY IM Equity | ITALY | Industrial | 12,356 | 3,849 | 1,962 | 0,159 | -0,563 | ** | **0,020** | |
| 28 | SAIPEM SPA | SPM IM Equity | ITALY | Energy | 28,707 | 61,638 | 7,851 | 0,273 | -0,577 | ** | -0,793 | ** |
| 29 | SNAM SPA | SRG IM Equity | ITALY | Utilities | 3,500 | 0,081 | 0,284 | 0,081 | **0,554** | | -0,351 | ** |
| 30 | STMICROELECTRONICS NV | STM IM Equity | ITALY | Technology | 5,804 | 1,919 | 1,385 | 0,239 | **0,635** | ** | -0,141 | |
| 31 | TELECOM ITALIA SPA | TIT IM Equity | ITALY | Communications | 0,951 | 0,018 | 0,135 | 0,142 | -0,277 | ** | -0,775 | ** |
| 32 | TENARIS SA | TEN IM Equity | ITALY | Industrial | 13,822 | 7,800 | 2,793 | 0,202 | -0,782 | ** | -0,282 | ** |
| 33 | TERNA SPA | TRN IM Equity | ITALY | Utilities | 2,870 | 0,091 | 0,302 | 0,105 | -0,061 | | -0,979 | ** |
| 34 | TOD'S SPA | TOD IM Equity | ITALY | Consumer, Cyclical | 64,372 | 338,238 | 18,391 | 0,286 | -0,255 | ** | -1,082 | ** |
| 35 | UNICREDIT SPA | UCG IM Equity | ITALY | Financial | 9,244 | 19,251 | 4,388 | **0,475** | -0,164 | ** | -1,363 | ** |
| 36 | UBI BANCA SCPA | UBI IM Equity | ITALY | Financial | 6,076 | 7,258 | 2,694 | **0,443** | -0,058 | | -1,501 | ** |

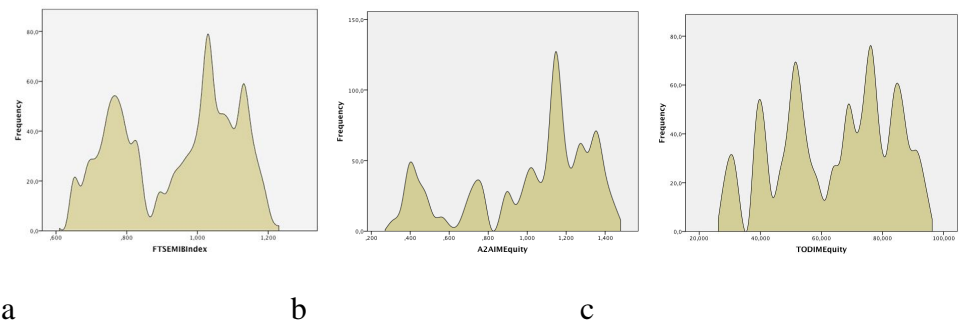

a          b          c

Figure 4: Frequency distributions of the price for some sample assets. (a) MIB Index; (b) A2A spa; (c) Tod's.

To measure the systematic risk, we considered different market proxies and we tested for the robustness of the results. In particular, we considered the Financial Time Stock Exchange Milano Indice Borsa (FTSE MIB) index, Interest Rate Swap (IRS) index at 5, 10 and 20 years (ILSW), Euro OverNight Index Average (Eonia) at 5, 10 and 20 years (EUSWE) and the EURO Inter Bank Offered Rate (Euribor) index at 12 months (EURO12). Only the results using the FTSE MIB index are reported here (see Table 1). Similar results were obtained using the other proxies. The analysis of market risk starts with the calculation of the joint second moment (covariance and correlation) of each asset price with the FTSE MIB and then moves to analyze the joint third moment characteristics (the standardized co-skewness, expressed by the parameters $\lambda_{2,1}$ and $\lambda_{1,2}$) and the joint fourth moment characteristics (the standardized co-kurtosis expressed by the parameters $\lambda_{1,3}$, $\lambda_{2,2}$ and $\lambda_{3,1}$). All these results are reported in Table 2. As expected there is a high and significant correlation between the two co-skewness measures and between the three co-kurtosis parameters. Figure 5 displays the scatter diagram of each asset with respect to the *FTSE MIB* index. Generally speaking observing Figure 5 it is evident that there are strong regularities in the data, but that they depart dramatically from the linearity paradigm. As a consequence, a simple correlation analysis is highly unsatisfactory if one wants to explain the complex links of dependence between each asset and the market. In fact, along with evidences of positive correlation, almost all graphs display marked departure from bivariate normality[3]. In particular there is a dominant feature

---

[3] Given the marked graphical characteristics and given the violation already tested in the univariate marginals, a formal tests of multinormality is not undertaken here. For details on

of strong positive lepto co-kurtosis which is evident from butterfly-shaped scatters and from a greater concentration of points in the tails of the joint bivariate distribution (that is in the top right and bottom left area of the graph).

There are only few remarkable exceptions to this characteristic which refer to 7 companies (namely Campari, Impregilo, Lottomatica, Luxottica, Pirelli, Saipem and Tod's; see scatters 12, 18, 20, 21, 26, 28 and 34 in Figure 5) which display, in contrast, evidences of a negative lepto-cokurtosis (that is, a higher concentration in the top left and bottom right area of the graph).

In the same Table 2, we also report the slope coefficients (representing the measures of systematic risk) calculated using the traditional *CAPM* expression based on the Least Squares method ($b_{LS}$) and the slope calculated with the alternative Least Quartic technique illustrated in Section 3 ($b_{LQ}$).

Generally speaking the sign of the $b_{LS}$ and $b_{LQ}$ are the same with few remarkable exceptions related to the seven assets which, in the previous analysis, were characterized by negative lepto-cokurtosis. In these seven cases, even if the assets' price appears to be negatively correlated with the risk free asset, they display a positive relationship in the tail. This feature is evident from the graphs where a dominant negative correlation is evident, but it is convoluted with a positive association of low and high tails with the *FSTE MIB* index.

*Table 2: Summary measures of co-skewness and co-kurtosis and slope evaluated through the LS criterion and the LQ methods. The last column reports the relative percentage difference between*

---

such tests see, e. g. Mardia, 1986

*the two slopes $_{\Delta b_{LQ}}$. Positive differences between the two systematic risk measures are highlighted in bold.*

| | Name | code | corr | std coskweness | | std cokurtosis | | | $b_{LS}$ | $b_{LQ}$ | $\Delta b_{LQ}$ |
|---|---|---|---|---|---|---|---|---|---|---|---|
| | | | | $\lambda_{1,2}$ | $\lambda_{2,1}$ | $\lambda_{1,3}$ | $\lambda_{2,2}$ | $\lambda_{3,1}$ | | | |
| | FTSE MIB INDEX | FTSEMIB Index | 1 | | 1 | 1 | 1 | 1 | | | |
| 1 | A2A SPA | A2A IM Equity | 0,818 | 0,807 | 0,827 | 0,821 | 0,814 | 0,793 | 1,625 | 1,051 | -35,32% |
| **2** | **ANSALDO STS SPA** | *STS IM Equity* | *0,851* | *0,849* | *0,842* | *0,843* | *0,826* | *0,844* | *5,998* | *7,608* | *26,85%* |
| 3 | **ASSICURAZIONI G** | *G IM Equity* | *0,941* | *0,937* | *0,929* | *0,931* | *0,908* | *0,930* | *14,936* | *15,568* | *4,23%* |
| **4** | **ATLANTIA SPA** | *ATL IM Equity* | *0,907* | *0,908* | *0,909* | *0,913* | *0,906* | *0,904* | *10,862* | *13,439* | *23,72%* |
| 5 | **AUTOGRILL SPA** | *AGL IM Equity* | *0,555* | *0,555* | *0,579* | *0,578* | *0,588* | *0,552* | *5,365* | *8,246* | *53,69%* |
| **6** | **AZIMUT HOLDING** | *AZM IM Equity* | *0,427* | *0,452* | *0,418* | *0,446* | *0,401* | *0,476* | *3,988* | *7,859* | *97,08%* |
| 7 | BANCA MONTE DE | BMPS IM Equity | 0,840 | 0,835 | 0,791 | 0,793 | 0,724 | 0,826 | 1,715 | 0,759 | -55,74% |
| **8** | BANCA POPOL EMI | BPE IM Equity | 0,877 | 0,867 | 0,882 | 0,877 | 0,872 | 0,853 | 11,371 | 8,223 | -27,68% |
| 9 | BANCA POPOLARE | PMI IM Equity | 0,797 | 0,798 | 0,755 | 0,764 | 0,711 | 0,796 | 2,373 | 1,053 | -55,65% |
| 10 | BANCO POPOLARE | BP IM Equity | 0,811 | 0,808 | 0,752 | 0,757 | 0,689 | 0,801 | 6,219 | 2,841 | -54,32% |
| 11 | **BUZZI UNICEM SPA** | *BZU IM Equity* | *0,755* | *0,760* | *0,744* | *0,754* | *0,726* | *0,764* | *7,525* | *9,371* | *24,53%* |
| 12 | DAVIDE CAMPARI-N | CPR IM Equity | -0,438 | -0,433 | -0,507 | -0,503 | -0,551 | -0,426 | -3,066 | 3,828 | 224,86% |
| 13 | DIASORIN SPA | *DIA IM Equity* | *0,379* | *0,373* | *0,390* | *0,381* | *0,396* | *0,363* | *13,183* | *25,954* | *96,88%* |
| 14 | ENEL SPA | *ENEL IM Equity* | *0,888* | *0,877* | *0,882* | *0,874* | *0,865* | *0,862* | *3,461* | *3,755* | *8,49%* |
| 15 | ENI SPA | *ENI IM Equity* | *0,422* | *0,428* | *0,417* | *0,425* | *0,412* | *0,434* | *3,127* | *15,818* | *405,85%* |
| **16** | FIAT SPA | *F IM Equity* | *0,370* | *0,367* | *0,364* | *0,358* | *0,350* | *0,362* | *3,135* | *4,802* | *53,17%* |
| 17 | FINMECCANICA SPA | FNC IM Equity | 0,839 | 0,833 | 0,812 | 0,811 | 0,768 | 0,822 | 15,889 | 8,314 | -47,67% |
| **18** | IMPREGILO SPA | IPG IM Equity | -0,199 | -0,170 | -0,254 | -0,224 | -0,300 | -0,140 | -0,533 | 2,745 | 615,27% |
| 19 | INTESA SANPAOLO | ISP IM Equity | 0,929 | 0,928 | 0,908 | 0,915 | 0,874 | 0,923 | 3,387 | 0,495 | -85,40% |
| **20** | LOTTOMATICA GRC | LTO IM Equity | -0,074 | -0,059 | -0,078 | -0,064 | -0,084 | -0,044 | -0,900 | 14,254 | 1684,37% |
| 21 | LUXOTTICA GROUP | LUX IM Equity | -0,368 | -0,355 | -0,442 | -0,429 | -0,493 | -0,340 | -10,528 | 22,853 | 317,06% |
| **22** | **MEDIASET SPA** | *MS IM Equity* | *0,907* | *0,900* | *0,884* | *0,884* | *0,836* | *0,889* | *8,154* | *39,777* | *387,85%* |
| 23 | MEDIOBANCA SPA | MB IM Equity | 0,889 | 0,883 | 0,880 | 0,881 | 0,852 | 0,873 | 9,614 | 6,934 | -27,87% |
| **24** | **MEDIOLANUM SPA** | *MED IM Equity* | *0,798* | *0,809* | *0,786* | *0,802* | *0,768* | *0,818* | *3,056* | *3,594* | *17,60%* |
| 25 | **PARMALAT SPA** | *PLT IM Equity* | *0,628* | *0,625* | *0,600* | *0,596* | *0,561* | *0,617* | *1,190* | *1,951* | *63,92%* |
| **26** | PIRELLI & C. | PC IM Equity | -0,449 | -0,439 | -0,532 | -0,523 | -0,571 | -0,427 | -5,587 | 6,164 | 210,33% |
| 27 | **PRYSMIAN SPA** | *PRY IM Equity* | *0,458* | *0,463* | *0,457* | *0,463* | *0,452* | *0,467* | *5,586* | *13,158* | *135,53%* |
| **28** | SAIPEM SPA | SPM IM Equity | -0,213 | -0,200 | -0,248 | -0,245 | -0,265 | -0,202 | -10,381 | 29,481 | 383,98% |
| 29 | **SNAM SPA** | *SRG IM Equity* | *0,339* | *0,330* | *0,344* | *0,335* | *0,348* | *0,320* | *0,599* | *3,616* | *503,58%* |
| **30** | STMICROELECTRON | STM IM Equity | 0,739 | 0,735 | 0,692 | 0,690 | 0,637 | 0,726 | 6,368 | 6,349 | -0,31% |
| 31 | **TELECOM ITALIA S** | *TIT IM Equity* | *0,900* | *0,896* | *0,896* | *0,896* | *0,886* | *0,889* | *0,755* | *1,001* | *32,62%* |
| **32** | **TENARIS SA** | *TEN IM Equity* | *0,281* | *0,286* | *0,290* | *0,294* | *0,299* | *0,288* | *4,883* | *14,145* | *189,70%* |
| 33 | **TERNA SPA** | *TRN IM Equity* | *0,458* | *0,454* | *0,470* | *0,465* | *0,479* | *0,446* | *0,860* | *29,438* | *3324,78%* |
| **34** | TOD'S SPA | TOD IM Equity | -0,279 | -0,275 | -0,301 | -0,299 | -0,303 | -0,270 | -31,906 | 67,014 | 310,04% |
| 35 | UNICREDIT SPA | UCG IM Equity | 0,952 | 0,951 | 0,928 | 0,937 | 0,878 | 0,946 | 25,982 | 9,944 | -61,73% |
| **36** | UBI BANCA SCPA | UBI IM Equity | 0,804 | 0,800 | 0,758 | 0,759 | 0,709 | 0,792 | 13,474 | 6,710 | -50,20% |

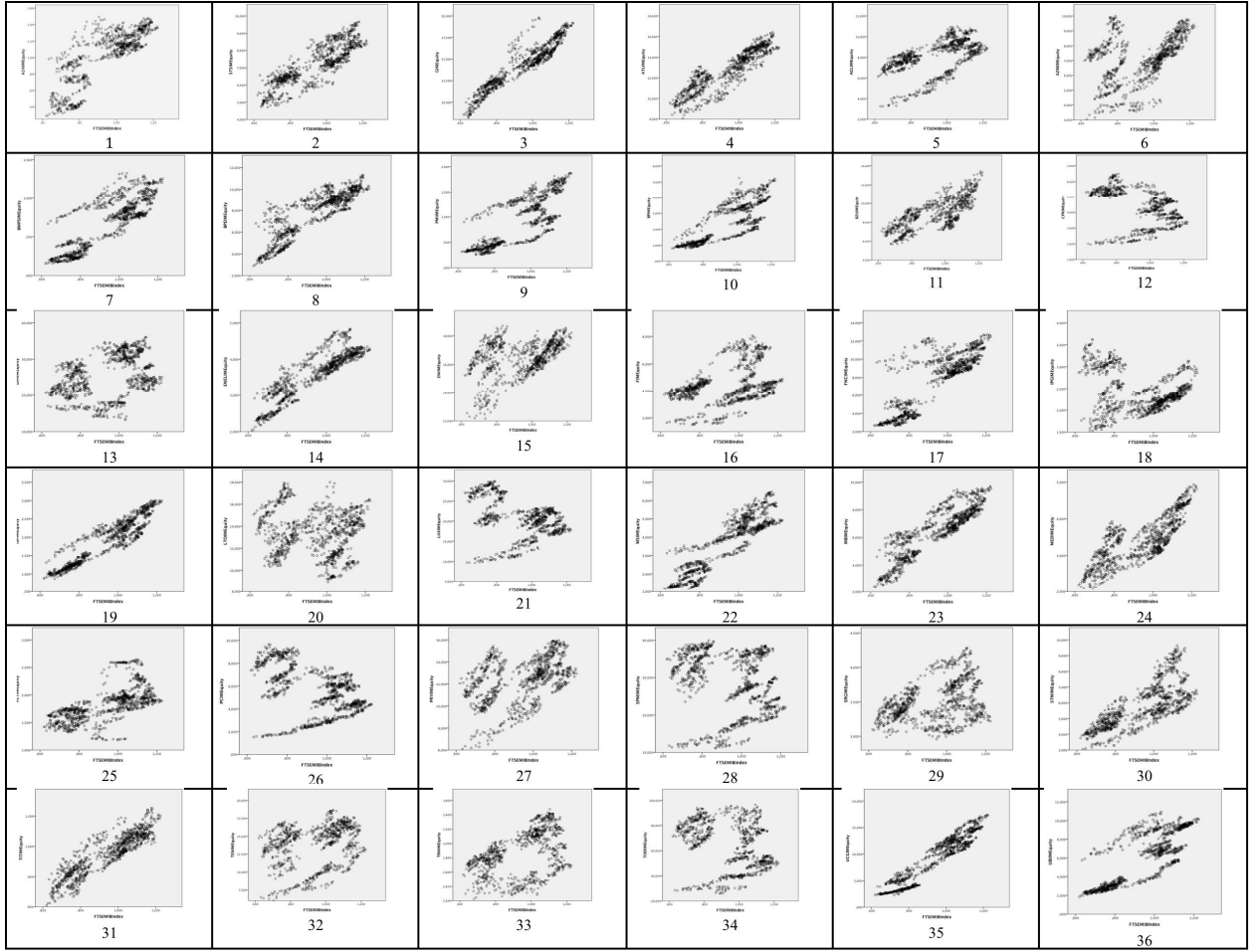

*Figure 5: Scatter diagram of the price of the FTSE MIB (horizontal axes) and each asset price (vertical axes). The numbers refer to the codes reported in Table 1 and 2.*

Table 2 also reports the relative difference between the traditional *CAPM* market risk based on the LS criterion and the risk evaluated with reference to the tail behavior, calculated as $\Delta b_{LQ} = \frac{b_{LQ} - b_{LS}}{b_{LS}} \times 100$. A positive (negative) value of $\Delta b_{LQ}$ thus indicates that we may incur into an underestimation (overestimation) of the systematic risk if we use the *Least Squares* instead of the *Least Quartic* criterion thus neglecting the joint tail information. In most cases the systematic risk is underestimated by the *LS* criterion with respect to the *LQ* criterion (25 cases out of 36). The cases of largest underestimation are those referring to Campari ENI, Impregilo, Lottomatica, Luxottica, Mediaset, Pirelli, Saipem, SNAM, Terna and TOD's. In particular, the two highest cases of underestimation occur in the case Lottomatica and Terna. Lottomatica is the case of a very low linear correlation with the risk-free asset (0.074), but with very strong tail dependence (see the butterfly-shaped scatter in graph 20 of Figure 5). In contrast, in the case of Terna, the correlation with FTSE MIB is positive (0.485), but the joint distribution departs dramatically from the Gaussian paradigm with all points in the scatter diagram that are arranged in a ring and with no points around the mean (see graph 33 in Figure 5) resulting in a very peculiar case of co-kurtosis which increases dramatically market risk. It is useful to recall that a ring-shaped distribution of points in the scatter diagram implies that when the FTSE MIB index presents a price around its mean the Terna price is far from its mean (and vice-versa) thus dramatically increasing the systematic risk of the asset.

Finally Table 3 reports the two ranking of the top ten companies in terms of systematic risk measure obtained using the two interpolation criteria (*LS* and *LQ*). It is evident a dramatic change in the evaluation of the systematic risk with only three companies being present in the top ten of both rankings (Assicurazioni Generali, Diasorin and Mediaset).

*Table 3: Ranking of the top 10 risky asset in the sample considered using the LS and the LQ criterion*

| ranking | Company name | $b_{LS}$ | ranking | Company name | $b_{LQ}$ |
|---|---|---|---|---|---|
| 1 | UNICREDIT SPA | 25,9816463 | 1 | TOD'S SPA | 67,0144 |
| 2 | FINMECCANICA SPA | 15,888672 | 2 | MEDIASET SPA | 39,7774 |
| 3 | ASSICURAZIONI GENERALI | 14,9358075 | 3 | SAIPEM SPA | 29,4812 |
| 4 | UBI BANCA SCPA | 13,4742741 | 4 | TERNA SPA | 29,4379 |
| 5 | DIASORIN SPA | 13,1826853 | 5 | DIASORIN SPA | 25,9542 |
| 6 | BANCA POPOL EMILIA ROMAGNA | 11,3714185 | 6 | LUXOTTICA GROUP SPA | 22,8526 |
| 7 | ATLANTIA SPA | 10,8624925 | 7 | ENI SPA | 15,8178 |
| 8 | MEDIOBANCA SPA | 9,61355394 | 8 | ASSICURAZIONI GENERALI | 15,5683 |
| 9 | MEDIASET SPA | 8,15356662 | 9 | LOTTOMATICA GROUP SPA | 14,2543 |
| 10 | BUZZI UNICEM SPA | 7,52540458 | 10 | TENARIS SA | 14,1451 |

## 5. CONCLUSION

Considerations related to the third and fourth moments characteristics within a regression framework are extremely relevant in many applied fields and particularly in market risk analysis of financial markets. In this paper we present a discussion on the statistical nature of co-skewness and co-kurtosis and we introduced a new criterion to estimate a linear regression model parameters, based on a quartic loss function. In quantitative finance this criterion could be exploited to evaluate market risk within the framework of the *CAPM* by taking into account third and fourth moments characteristics of the asset price distribution. The potential of the method is illustrated with reference to an example based on some Italian stock market price data. The empirical analysis, based on the Least Quartic estimation of the slope coefficient, adds insights in market analysis and helps in identifying potentially risky assets whose extreme behavior is strongly dependent on the market behavior.

A number of generalizations could be taken into consideration in the future to expand the results presented here. A first extension is in the direction of the statistical estimation. In this respect, future studies in this field could be directed towards the analysis of regressions estimation within a Maximum

Likelihood framework. The importance of higher-order considerations in likelihood-based estimation was recently pointed out by Holly *et al.* (2011) who considered fourth-order maximum likelihood techniques based on an exponential family specification of the likelihood function. Following their suggestion one could specify a bivariate joint distribution, say $f_{x_i y_i}(x_i, y_i)$ by using a bivariate Pearson's curve (as suggested e. g. by Premaratne and Bera, 2001) or a bivariate exponential family curve or a mixture distribution. The full likelihood could then be derived as the product of the bivariate marginals $l(\theta) = \prod_{i=1}^{n} f_{x_i y_i}(x_i, y_i)$ (θ being a set of parameters). In the *CAPM* application that we considered in this paper, this approach would lead to the estimation of parameters that express more thoroughly the complex relationships of dependence between each asset and the market thus representing alternative measurements of the asset's market risk.

In terms of the financial applications of the least quartic regression, a second possible extension is in the field of portfolio optimization. In fact, the traditional portfolio theory (Markovitz, 1959) is based on the idea of finding the optimal weights of a portfolio by minimizing the portfolio variance which, in turn can be expressed as a linear combination of each asset's variance and of the covariance between all pairs of assets in the portfolio. However, based on an implicit assumption of joint normality of assets' returns, this strategy neglects the effects of third and fourth order moments of the distributions. The importance of higher moments in optimal portfolio allocation was considered for instance by Jondeau and Rockinger (2006). In this respect, following the idea on which this paper is based, one may think of minimizing the fourth moment of the portfolio instead of its variance. In this way the optimal weights could be derived as a function of the kurtosis and of the pairwise co-kurtosis of the assets thus deriving optimal portfolios that adequately account for the extreme behavior of the assets price.

Finally it should be noted that a part of the financial econometrics literature (e. g. Brooks *et al.*, 2005; Dubauskus and Teresiene, 2005) has considered models based on conditional higher moments which have essentially extended the ARCH formulation of Engle (1982). In contrast the criterion adopted in this paper is based on unconditional moments and co-moments. Furthermore our approach also differs from the one adopted by Conrad et al. (2013) which is based on *risk-neutral moments*. Extending our approach to both conditional and risk-neutral moments represents a further area of possible development of the present contribution.

REFERENCES

Arbia, G. (2003) Bivariate Value-at-risk, *Statistica*, 62, 2, 231-247.

Bakshi, G. , Kapadia, N. and Madan, D. (2003) Stock return characteristics, skew laws, and the differential pricing individual equity options, *Review of Financial Studies,* 16, 101-143.

Barberis, N. and Huang, M. (2008) Stock as lotteries: The implications of probability weighting for security prices, *American Economic Review,* 98, 2066-2100.

Bargès, M., Cossette, H. and Marceau, E. (2009). TVaR-based capital allocation with copulas. *Insurance: Mathematics and Economics*, 45, 348-361.

Barone-Adesi, G. (1985) Arbitrage equilibrium with skewed asset returns, *Journal of Financial and Quantitative Analysis*, 20, 3, 299-313.

Barone-Adesi, G., and Talwar, P. (1983) Market Models and Heteroscedasticity of Residual Security Returns, *Journal of Business & Economic Statistics*, 1, 163–168.

Barone-Adesi, G., Gagliardini, P. and Urga, G. (2004) Testing Asset Pricing anomalies with Coskewness, *Journal of Business and Economic Statistics*, 22, 474-485.

Boyer, B., Mitton, T. and Vorkink, K. (2010) Expected idiosyncratic skewness, *Review of Financial Studies,* 23, 169-202.

Brooks, C. Burhe, S. P. Hervai, S. and Pertsand, G. (2005), Autoregressive conditional kurtosis, *Journal of Financial Econometrics*, 3, 399-421.

Brunnermeier, M. K. Gollier, C. and Parker, J. A. (2007) Optimal beliefs, asset prices, and the preference for skewed returns, *American Economic Review,* 97, 159-165.

Christie-David, R., Chaudry, M. (2001) Coskewness and cokurtosis in futures markets, *Journal of Empirical Finance*, 8, 55-8.

Conrad, J., Dittmar, R. F. and G, E. (2013) Ex ante skewness and expected stock returns, *Journal of Finance,* 68, 1, 85-124.

Dittmar, R. (1999). Nonlinear pricing kernels, kurtosis preference, and evidence from the cross-section of equity returns, *Journal of Finance*, 1, 369-403.

Dittmar, R. (2002) Nonlinear pricing kernels, kurtosis preference, and evidence from the cross section of equity returns, *Journal of Finance,* 57, 369-403.

Dubauskus, G. and Teresiene, D. (2005), Autoregressive conditional skewness and kurtosis and Jarque Bera statistics in Lithuanian stock market measurement, *ISSN Engineering Economics*, 5, 45, 19-24.

Fama, E. and French, K. (1992) The cross section of expected returns, *Journal of Finance*, 59, 427-465.

Fang H. and Lai, T.-Y-. (1997) Cokurtosis and capital asset pricing, *The Financial Review*, 32-293-307.

Harvey, C. and Siddique, A. (1999) Autoregressive conditional skewness in asset pricing tests, *Journal of Financial and Quantitative Analysis*, 34, 465-487.

Harvey, C. and Siddique, A. (2000) Conditional skewness in asset pricing tests, *Journal of Finance*, 3, 1263-1295.

Green, T. C. and Hwang, B-H. (2012) IPO as lotteries: Expected skewness and first-day returns, *Management Science* 58, 432-444.

Holly A., Monfort A. & Rockinger M. (2011) Fourth order pseudo maximum likelihood methods. *Journal of Econometrics*, *162*, 2, 278-293

Hwang, S., Satchell, S. (1999) Modeling emerging market risk premia using higher moments, *International Journal of Finance & Economics*, Vol. 4, 271-296.

Ingersoll, J. (1975) Multidimensional security pricing, *Journal of Financial and Quantitative Analysis*, 10, 85-798.

Jacobson, N. (2009) *Basic algebra*, volume 1, (second edition), Dover

Jondeau, E. and Rockinger, M. (2006) Optimal portfolio allocation under higher moments, *European Financial Management*, 12, 1, 29-55.

Kendall M. G., Stuart, A. and Ord, J. K. (1983) *The advanced theory of statistics*, Vol. 1, Griffin, London.

Kotz, S., Balakrishnan, N. and Johnson, L. (2000) *Continuous multivariate distributions. Models and applications*, John Wiley and sons.

Kraus, A., Litzenberger, R. (1976) Skewness preference and the valuation of risky assets, *Journal of Finance*, 31, 1085-1100.

Kraus, A., Litzenberger, R. (1983) On the distributional conditions foe a consumption oriented three moment CAPM, *Journal of Finance,* 38, 1381-1391.

Krämer, W. and Runde, R. (2000) Peaks or tails: what distinguishes financial data ?, *Empirical Economics*, 25, 665-671.

Lintner, J. (1965) Security Prices, Risk, and Maximal Gains from Diversification, *Journal of Finance*, 20, 587-615

Mardia, K. V. (1986) Mardia's test of multinormality, in N. L. Johnson, S. Kotz and C. Read. (eds.) *Encyclopedia of statistics*, 5, Wiley, New York, 217-221.

Mitton, T. and Vorkink, K. (2007) Equilibrium underdiversification and the preference for skewness, *Review of Financial Studies,* 20, 1255-1288.

Mossin, J. (1966) Equilibrium in a capital asset market, *Econometrica*, 768-783.

Markovitz, H. (1952) Portfolio selection, *The Journal of Finance*, 7, 1, 77-91.

McNeil, A. J., Frey, R. (2000) Estimation of tail-related risk measures for heteroscedastic financial time series: an extreme value approach, *Journal of Empirical Finance*, 7, 3–4, 271–300

Premaratne, G. and Bera, A. (2001) Modeling asymmetry and excess kurtosis in stock return data, Mimeo, Department of Economics, University of Illinois at Urbana Champain.

Ranaldo, A., L. Favre (2005): Hedge Fund Performance & Higher-Moment Market Models. *Journal of Alternative Investments*, 8, 3, 37-51.

Rubinstein, M. E. (1994) Implied binomial trees, *Journal of Finance*, 52, 35-55.

Sears, S. and Wei, K. C. J. (1988) The structure of skewness preferences in asset pricing model with higher moments: an empirical test. *Financial Review*, 23, 25-38.

Sen, P. K. (1968) Estimates of the regression coefficient based on Kendall's tau, *Journal of the American Statistical Association,* 63, 1379–89,

Sharpe, W. F. (1964) Capital Asset Prices: a Theory of Market Equilibrium under Conditions of Risk, *Journal of Finance,* 19, 425-442.

Theil, H. (1950) A rank-invariant method of linear and polynomial regression analysis. I, II, III, *Nederl. Akad. Wetensch., Proc.* 53, 386–92, 521–5, 1397–412.